\documentclass[final,3p,times,twocolumn,fleqn]{elsarticle}

\usepackage{amssymb}

\usepackage{amsmath}

\journal{Chaos, Solitons and Fractals}

\begin{document}

\begin{frontmatter}



\title{Polarized opinion states in static networks driven by limited information horizons}

\author[gotham,matCond]{H. P\'erez-Mart\'inez\corref{cor}}
\cortext[cor]{Corresponding author}
\ead{h.perez@unizar.es}
\author[gotham,fisTeor]{F. Bauz\'a Mingueza}
\author[gulb,gotham]{D. Soriano-Pa\~nos}
\author[gotham,matCond]{J. G\'omez-Garde\~nes}
\author[gotham,matCond]{L.M. Flor\'ia}

\affiliation[gotham]{
    organization={GOTHAM lab, Institute for Biocomputation and Physics of Complex Systems (BIFI), University of Zaragoza},
    postcode={50018},
    city={Zaragoza},
    country={Spain}
    }

\affiliation[matCond]{
    organization={Department of Condensed Matter Physics, University of Zaragoza},
    postcode={50009}, 
    city={Zaragoza},
    country={Spain}
    }

\affiliation[fisTeor]{
    organization={Department of Theoretical Physics, University of Zaragoza},
    postcode={50009}, 
    city={Zaragoza},
    country={Spain}
    }
    
\affiliation[gulb]{
    organization={Institute Gulbenkian of Science (IGC)},
    postcode={2780-156},
    city={Oeiras},
    country={Portugal}
    }

\begin{abstract}
The widespread emergence of opinion polarization is often attributed to the rise of social media and the internet. These platforms can promote selective exposure, leading to the formation of echo chambers where individuals are only exposed to viewpoints that reinforce their existing beliefs. However, experimental evidence shows that exposure to opposing views through cross-cutting ties is common in both online and offline social contexts, which frequently involve long-standing personal relationships. To account for these facts, we have developed an opinion model that applies to static contact structures. In this model, an agent's influence over their neighbors depends on the similarity of their opinions. Our findings suggest that polarization can indeed emerge in such static structures and, driven by an increased narrow-mindedness, even in presence of non-negligible cross-cutting ties.  Interestingly, the polarized opinion distributions generated by our model closely resemble those obtained in surveys about highly polarized issues. This has allowed us to categorize various issues based on their controversial nature, sheding light on the factors that contribute to opinion polarization.
\end{abstract}



 


\end{frontmatter}


\section{Introduction}
\label{sec:intro}

Opinion polarization is on the rise on modern societies~\cite{baldassarri2007dynamics, baldassarri2008constraint, gentzkow2016polarization, abramowitz2019racial}, and its presence leads to significant impacts both in human behavior and in social relationships. A clear example of such increasing influence of polarization can be found in the political sphere. For instance, in~\cite{wells2017conversation} a decrease in cross-cutting political talk has been recorded following a particularly harsh confrontation between the governor of Wisconsin in 2011 and public-sector employees. Moreover, different studies suggest that severe polarization leads to legislative gridlock, democratic backsliding, political collapse~\cite{mccoy2018democracy}, and the decrease of cross-party cooperation~\cite{iyengar2018partisan}. Polarization also affects health issues, biasing drug prescription~\cite{hersh2016physicians} on highly politicized issues, such as drug abuse or abortion, or compromising  vaccination campaigns due to strong in-group identification of non-vaccinated individuals. From a social perspective, polarization sparks discrimination~\cite{iyengar2015partisanship} and influences dating behavior~\cite{huber2017dating} to a degree comparable to discrimination based on race, hampering social gatherings. For instance, in~\cite{chen2018thanksgiving}, the authors estimate a loss of 34 million hours in family gatherings during Thanksgiving due to partisan effects derived from the US national elections in 2016.

Scholars have identified social media and the pervasive use of the internet as some of the factors contributing to this phenomenon~\cite{farrell2012internet, allcott2020media}. These platforms can promote selective exposure as they allow for the formation of echo chambers around polarized issues like gun control, abortion or vaccination~\cite{schmidt2018vaccination, cinelli2021echo} which can fuel radicalization processes~\cite{levy2021media}. Some recent studies performed over web-browsing data suggest that online media consumption is highly segregated in terms of political orientation and engagement~\cite{garimella2021media}, and that social media or web-search engines exacerbates this tendency~\cite{flaxman2016bubbles}.

From a complex systems perspective, research has been devoted to studying echo chambers and opinion segregation in social media, taking into account the dynamical aspect of online social networks.In this sense, two major approaches have been developed. The first one introduces contact networks gradually changing over time, mimicking the behavior of social network in platforms like Twitter. In these networks, links are created or broken either depending on opinion similarity~\cite{santos2021dynamical} or as a response to exposure to cross-cutting messages~\cite{sasahara2021unfollowing, tokita2021cascade}. The other main approach uses temporal networks in which the probability of interacting with any other agent of the system is proportional to the opinion similarity~\cite{baumann2020, baumann2021}. In both approaches, the major role of echo chambers and filter bubbles in the process of opinion formation of an agent is stressed.

In contrast, some studies suggest that the importance of echo chambers and selective exposure in social media on the polarization processes of modern societies might have been overstated~\cite{dubois2018echo, guess2018echo, benkler2018propaganda,zhuravskaya2020socialmedia}. On the one hand, some surveys have shown little connection between social media consumption and the increase of polarization~\cite{barbera2014polarization}, which becomes more relevant within those groups least likely to use the Internet~\cite{boxell2017demographic}. On the other hand, echo chambers could be less common than previously thought, as individuals rarely become isolated from opposing points of view. For example, in~\cite{huckfeldt2004ambivalence} survey-based data about the 2000 general election in the United States prove that more than half of the respondents reported cross-cutting political conversations. Moreover, recent studies with Reddit data~\cite{francisci2021reddit, monti2023reddit} actually show a preference of political interaction through cross-cutting ties.
Likewise, offline social contexts, such as the workplace~\cite{mutz2006workplace} or friendship circles~\cite{goel2010perceived}, usually feature cross-cutting interactions, which can be transferred to social media due to the close relationship between offline and online social interactions~\cite{bakshy2015facebook}.

Stable interaction patterns, inherent to daily-life social relationships and often featuring cross-cutting interactions~\cite{offer2018demanding}, thus remain a key actor in the emergence of polarization. Multiple opinion models applicable to static contact structures have been previously developed to study the phenomenon of opinion formation, including the Deffuant model~\cite{deffuant2000bounded} and the Hegselmann-Krause model~\cite{hegselmann2002bounded}. Both are well-known bounded confidence models, in which agents only interact if the distance between their opinions is smaller than a certain threshold. Both models assume a well-mixed population in their original formulations, but their study have been also successfully extended to static networks~\cite{meng2018networks}. In these works, the dynamics leads to either consensus, bipolarization or fragmentation of the opinion spectrum~\cite{lorenz2007review,lorenz2008meanfield}, in which agents usually cluster around a finite set of opinions, their number depending on the parameters of the system. Agents also become completely isolated from other points of view, as they lie outside their thresholds.

These models also have some drawbacks, as they consider that individuals give the same importance to all agents regardless of the distance between their opinions, provided that they lie inside their confidence threshold. To alleviate this assumption, some notable extensions have been proposed: the first one is the {\em Relative Agreement Model}~\cite{deffuant2002relative}, in which the thresholds can change during the opinion formation process, and the influence of each neighbor over an agent depends on the overlapping between their thresholds. The second one is the {\em Smooth Bounded Confidence Model}~\cite{deffuant2004smooth}, in which the influence is assumed to decay following a gaussian distribution of variable standard deviation. Regardless, both modifications remain conceptually bounded confidence models, and qualitatively reproduce the same final opinion distributions as the original works.

The former models fail to replicate what is observed in real world surveys about highly polarized issues, in which people do not follow a limited set of positions but rather display a wide range of opinions often resulting in a bimodal distribution~\cite{fiorina2008bimodal,ANES2016}. Moreover, they do not capture cross-cutting interactions that remain relevant in the context of daily life interaction~\cite{huckfeldt2004ambivalence}. These limitations warrant for an increased effort to develop models to better reproduce polarization in static networks. To fill this gap, here we propose a new model of opinion formation over static contact structures. Our model  allows for communication between disagreeing individuals and consider the homophily effect by which people tend to value more the opinions of like-minded individuals. Phenomenologically, our model overcomes the former limitations as polarized configurations with bimodal opinion distributions naturally arises from the dynamics.

The paper is organized as follows: in section~\ref{sec:model} we present the assumptions and the equations of the model. In section \ref{sec:fully} we analytically characterize  the proposed model in a fully connected graph, and in section~\ref{sec:network} we present the numerical results obtained by applying the model over different network structures. In section~\ref{sec:exp} we show how the model allows retrieving opinion distributions from experimental data and extract information about highly polarized issues. Finally, in section~\ref{sec:conc} we summarize the main conclusions of our work.

\section{The model}
\label{sec:model}

We consider a system of $N$ interacting agents, whose acquaintances do not change over time and are codified in an unweighted undirected adjacency matrix $A_{ij}$. Agents' opinions are characterized by a continuous variable $x_i(t)$; the sign $\sigma(x_i)$ represents the qualitative position of the agent over a certain issue (for or against), and the modulus $|x_i|$, her degree of radicalization.

For the opinion formation process, we choose to follow the approach in previous works~\cite{baumann2020, santos2021dynamical} which assumes that agent's opinions are influenced by their neighbors, and change over time following three basic principles: \textit{(i)} agents lose memory of their opinions with time, \textit{(ii)} they tend to align their opinion with their neighbors, and \textit{(iii)} they give more relevance to like-minded neighbors. However, contrary to those models, we assume that interaction patterns remain stable with time, thus implying a certain degree of cross-cutting interactions regardless of the agents' opinions, and in turn consider that individuals do not pay the same attention to everyone whom they interact with. Taking these mechanisms into account, the time evolution of the opinion for agent $i$ can be expressed as:
\begin{equation}
    \dot{x}_i(t) = -x_i(t) + K \sum_{j=1}^N A_{ij}w_{ij}(t) \tanh{x_j(t)} \; ,
    \label{eq:xChange}
\end{equation}

\noindent where $K$ is the \textit{social interaction strength}, which determines the strength that neighbors' opinions have over the agent's, and the weights $w_{ij}$ represent how valuable the view of neighbor $j$ is to agent $i$. We choose this function as:
\begin{equation}
    w_{ij}(t) = \frac{(|x_i-x_j|+\delta)^{-\beta}}{\sum_{l=1}^{N} A_{il}(|x_i-x_l|+\delta)^{-\beta}} \; ,
\end{equation}

\noindent where $\beta$ is the \textit{homophily parameter} (the higher, the more influential like-minded neighbors), and $\delta$ is a parameter included as a small noise, taken to be $\delta=0.002K$ throughout the paper. Note that weights change over time, fulfilling $\sum_j w_{ij} = 1$, and that neighbors' influence might not be symmetric, so that $w_{ij} \neq w_{ji}$ in general. Moreover, differently to the previously mentioned bounded-confidence models, our weights choice ensures that communication between disagreeing individuals remains relevant in the dynamics regardless of the distance in the opinion space.

The system develops on an $N$-dimensional phase-space formed by the opinions of all agents, and its behavior is defined by a system of $N$ differential equations whose asymptotic solutions give the opinions' stable equilibrium configurations. Although this system cannot be solved analytically in general, we can study its behavior in some particular cases. First, we consider a situation in which all agents hold the same opinion, $x_i = x \; \forall i$. In this situation, the $N$-equation system is reduced to a single equation $\dot{x} = -x + K\tanh{x}$. The expected opinion of all agents in the equilibrium $x^*$ is then given by:
\begin{equation}
    x^* = K\tanh{x^*}.
\end{equation}

\noindent For $K\leq1$, all agents reach $x = x^* = 0$ and the system arrives to a neutral consensus state. For $K>1$, all agents reach non-zero final opinions, ending up in a radicalized state. Therefore, there exists a second order transition in $K=1$ characterized by the order parameter $x^*$. Note that $x^*$ also corresponds to the maximum opinion value an agent can adopt given a $K$ value in the general scenario. Therefore, agents' opinions $x_i$ are bounded, fulfilling $x_i \in [-x_{\text{max}},x_{\text{max}}]$, with $x_{\text{max}}=x^*$. This result does not depend on the agent's degree due to the normalization of the weights.

\section{Fully-connected graph}
\label{sec:fully}

The homogeneity assumption is incapable of predicting the existence of polarization, defined as the coexistence of opposing views. In order to analytically characterize the possible emergence of polarization, we consider a fully-connected graph, so that $k_i=N-1 \; \forall i$, and perform a stability analysis of a polarized configuration. Let us assume that such polarized configuration corresponds to an equilibrium state of $n$ agents holding opinions $x^+>0$ and $m$ agents holding opinions $x^-<0$, with $n+m=N$, $n>2$, $m>3$. Values $x^+$, $x^-$ are given by the solution the system of equations obtained from Eq.~\ref{eq:xChange}:
\begin{subequations}
\begin{equation}
x^+ = K \frac{(n-1)\tanh{x^+} + m \left(\frac{\delta}{x^+-x^-+\delta}\right)^\beta \tanh{x^-}}{n-1 + m \left(\frac{\delta}{x^+-x^-+\delta}\right)^\beta} \; ,
\label{eq:x+}
\end{equation}
\begin{equation}
x^- = K \frac{n\left(\frac{\delta}{x^+-x^-+\delta}\right)^\beta \tanh{x^+} + (m-1) \tanh{x^-}}{n \left(\frac{\delta}{x^+-x^-+\delta}\right)^\beta + m-1} \; .
\label{eq:x-}
\end{equation}
\end{subequations}

\begin{figure}[t]
    \centering
    \includegraphics[width=.5\textwidth]{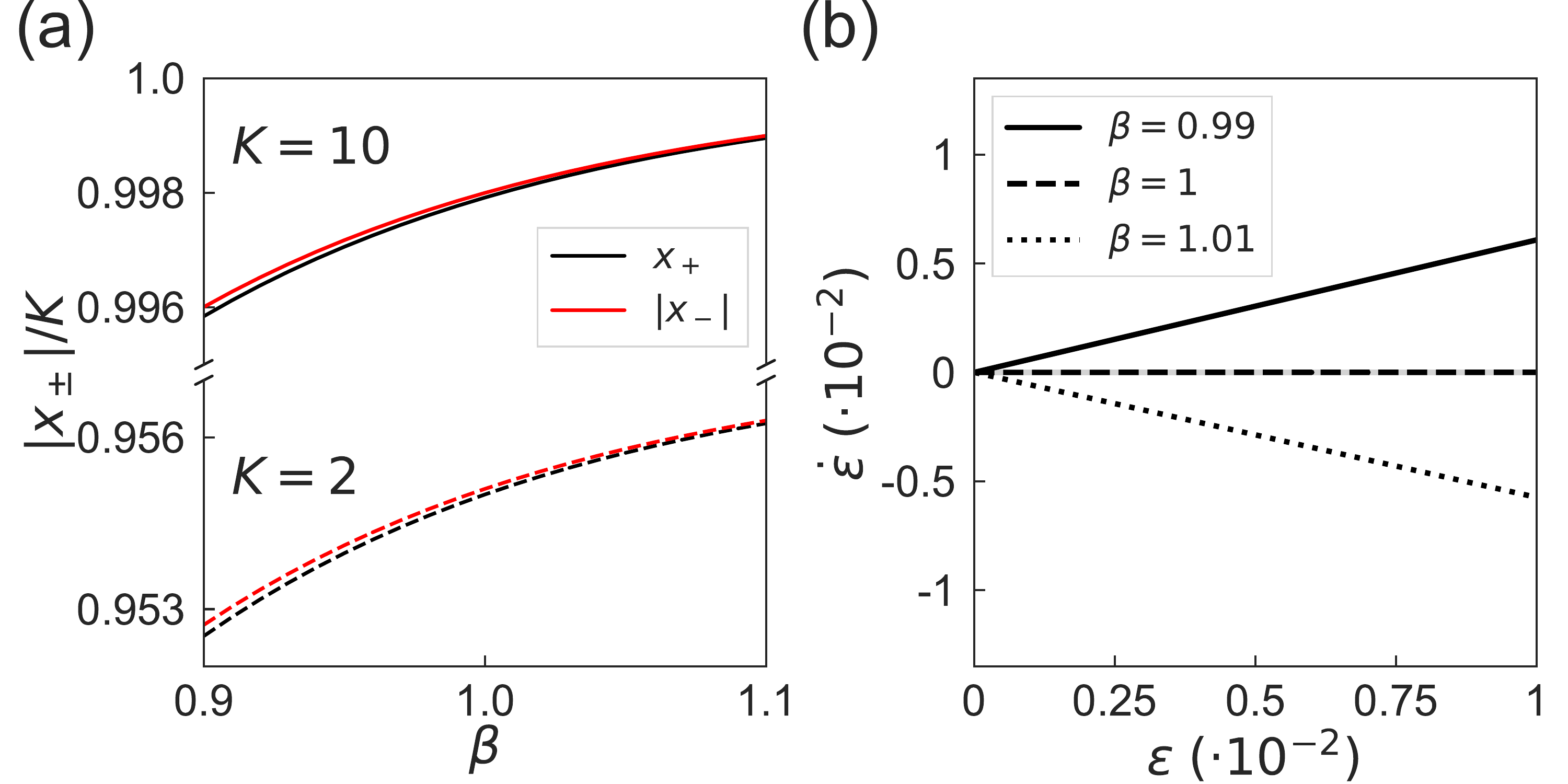}
    \caption{\label{fig:MF} (a) Values of $x_+$ (black), $x_-$ (red) obtained from Eqs.~\ref{eq:x+} and \ref{eq:x-}. Results are shown for $K=10$ (top, solid lines) and $K=2$ (bottom, dashed lines). (b) Values of $\dot{\epsilon}$ as a function of $\epsilon$ given by Eq.~\ref{eq:epsilon} for $K=10$. The perturbation grows below the transition ($\beta<1$) destabilizing the polarized states, and diminishes above the transition ($\beta>1$). For both panels we fix $m=51$, $n=50$, and $\delta=0.002K$.}
\end{figure}

This system can be solved numerically, and its behavior for different elections of the parameters is shown in Fig.~\ref{fig:MF}(a). The equilibrium values of $x_+$ and $x_-$ depend on the relative number of individuals with positive and negative opinions ($n$, $m$), and the values of parameters $\beta$ and $K$. For higher values of $\beta$, agents become increasingly isolated from opposing points of view, increasing $|x_{\pm}|$. Higher values of $K$ also affect the equilibrium opinions through $\delta$, resulting in increased radicalization.

We study the stability of the presented polarized configurations by adding a small perturbation $\epsilon$ to agent's $i$ opinion, so that $x_i = x^-+\epsilon$. Her opinion changes following the equation:
\begin{align}
    \dot{x}_i &= \dot{\epsilon}= -(x^- + \epsilon) + \nonumber \\ &+K \left[ \frac{(m-1)\tanh(x^-) + n\tanh(x^+)\left(\frac{\epsilon+\delta}{x^+-x^--\epsilon +\delta}\right)^{\beta}}{(m-1) + n \left(\frac{\epsilon+\delta}{x^+-x^--\epsilon + \delta}\right)^{\beta}} \right] \; .
\end{align}

\noindent Assuming $\epsilon<<\delta$, we can approximate:
\begin{align}
    &\left(  \frac{\delta+\epsilon}{x^+-x^--\epsilon + \delta} \right)^\beta \simeq \nonumber \\ & \left( \frac{\delta}{x^+-x^-+\delta} \right)^\beta + \beta \frac{x^+-x^-+2\delta}{(x^+-x^-+\delta)^2} \left( \frac{\delta}{x^+-x^-+\delta} \right)^{\beta-1} \epsilon \; .
\end{align}

\noindent Furthermore, considering Equation~\ref{eq:x-} and the approximation for small $\epsilon$:
\begin{equation}
    \frac{a\epsilon + b}{c\epsilon + d} \simeq \frac{b}{d} + \frac{ad-bc}{d^2}\epsilon \; ,
\end{equation}

\noindent we obtain:
\begin{equation}
    \dot{\epsilon} = \epsilon \left[-1 + A \left(\frac{\delta}{x^+-x^- + \delta}\right)^{\beta-1} \right] \; ,
    \label{eq:epsilon}
\end{equation}

\noindent where
\begin{equation}
    A = \frac{\beta Kn(m-1)(\tanh{x^+}-\tanh{x^-})\frac{x^+-x^- + 2\delta}{(x^+-x^- + \delta)^2}}{\left[m-1 + n \left(\frac{\delta}{x^+-x^-+\delta}\right)^\beta\right]^2} \; .
\end{equation}

\noindent The stability of the polarized system thus depends on the second term on the right hand side of Eq.~\ref{eq:epsilon}. On the one hand, the factor multiplying $A$ decreases with $\beta$, and is equal to $1$ when $\beta=1$. On the other hand, for large $K$ values, we can approximate $x^+ = -x^- \simeq K$, $\tanh(x^+) \simeq 1$, and, considering an evenly polarized system, we have $m \simeq n$. If $\delta$ is small enough, we find that $A \simeq \beta$, and thus, $A$ grows monotonically with $\beta$. Nevertheless, it can be seen in Fig.~\ref{fig:MF}(b) that the perturbation grows if $\beta>1$ ($\dot{\epsilon}>0$), and diminishes otherwise. Therefore, this kind of configurations becomes unstable for $\beta>1$.

\begin{figure}[t]
    \centering
    \includegraphics[width=.5\textwidth]{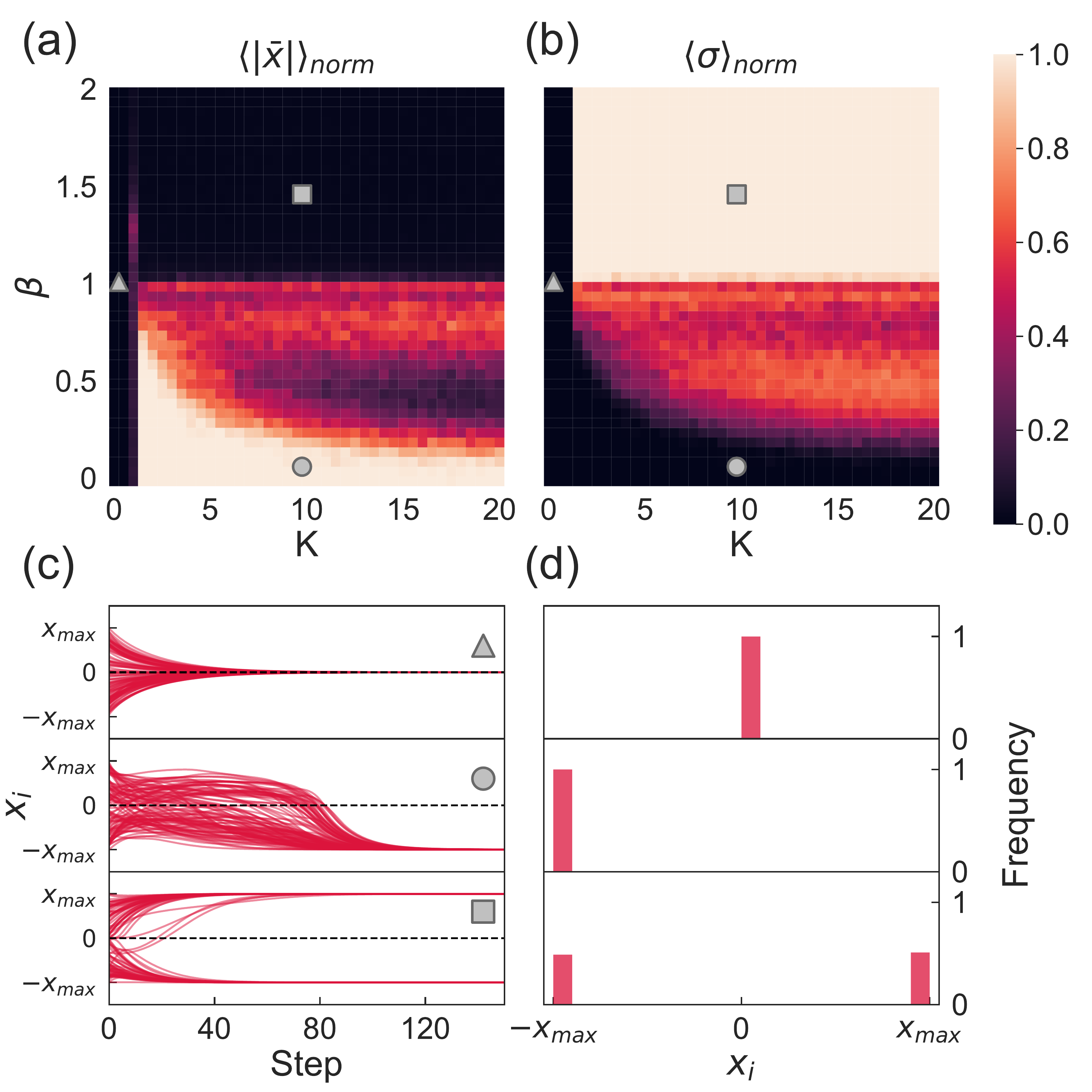}
    \caption{\label{fig:ER} (a) Average mean opinion $\langle |\bar{x}|\rangle_{norm}$ and (b) standard deviation $\langle \sigma\rangle_{norm}$ obtained from 100 independent realizations for each combination of parameters $(K, \beta)$ (color code). In each realization, initial opinions are randomly chosen on the interval $[-x_{\text{max}}, x_{\text{max}}]$. (c) Temporal evolutions of selected agents in the representative configurations: neutral consensus (top, $K=0.5$, $\beta=1$), radicalization (center, $K=10$, $\beta=0.1$) and polarization (bottom, $K=10$, $\beta=1.5$). (d) Histograms corresponding to the final configurations of panel (c). The underlying structure of contacts throughout the figure is an Erd\"os-R\'enyi network of $N=10^4$ nodes with mean degree $\langle k\rangle=10$. For the sake of clarity, only the trajectories of $1\%$ of the agents have been represented in panel (c).}
\end{figure}

\section{Numerical results on networked populations}
\label{sec:network}
In order to characterize the model's behavior in realistic environments, in which the agents usually do not know the opinion of each of the other components of the system, we consider static graphs with limited information horizons, and we solve the system of $N$ coupled equations using an explicit fourth-order Runge-Kutta method with $dt=0.1$, starting from uniformly distributed initial states within the range $[-x_{\text{max}}, x_{\text{max}}]$. To characterize a given equilibrium configuration, we use the absolute value of the mean opinion of the system $|\bar{x} |= |\sum_i x_i|/N$ and its standard deviation $\sigma$, given by $\sigma^2 = \sum_i (x_i-\bar{x})^2/N$. Both order parameters allow us to classify configurations as neutral consensus $(|\bar{x}| = 0$, $\sigma = 0)$, radicalized states $(|\bar{x}| \neq 0$, $\sigma \simeq 0)$ and polarized configurations $(|\bar{x}| \simeq 0$, $\sigma \neq 0)$. Furthermore, to classify intermediate situations in which $|\bar{x}| \neq 0$, $\sigma \neq 0$ we consider that a configuration in which $|\bar{x}|<\sigma$ is polarized, and radicalized otherwise.

\begin{figure}[t]
    \centering
    \includegraphics[width=.5\textwidth]{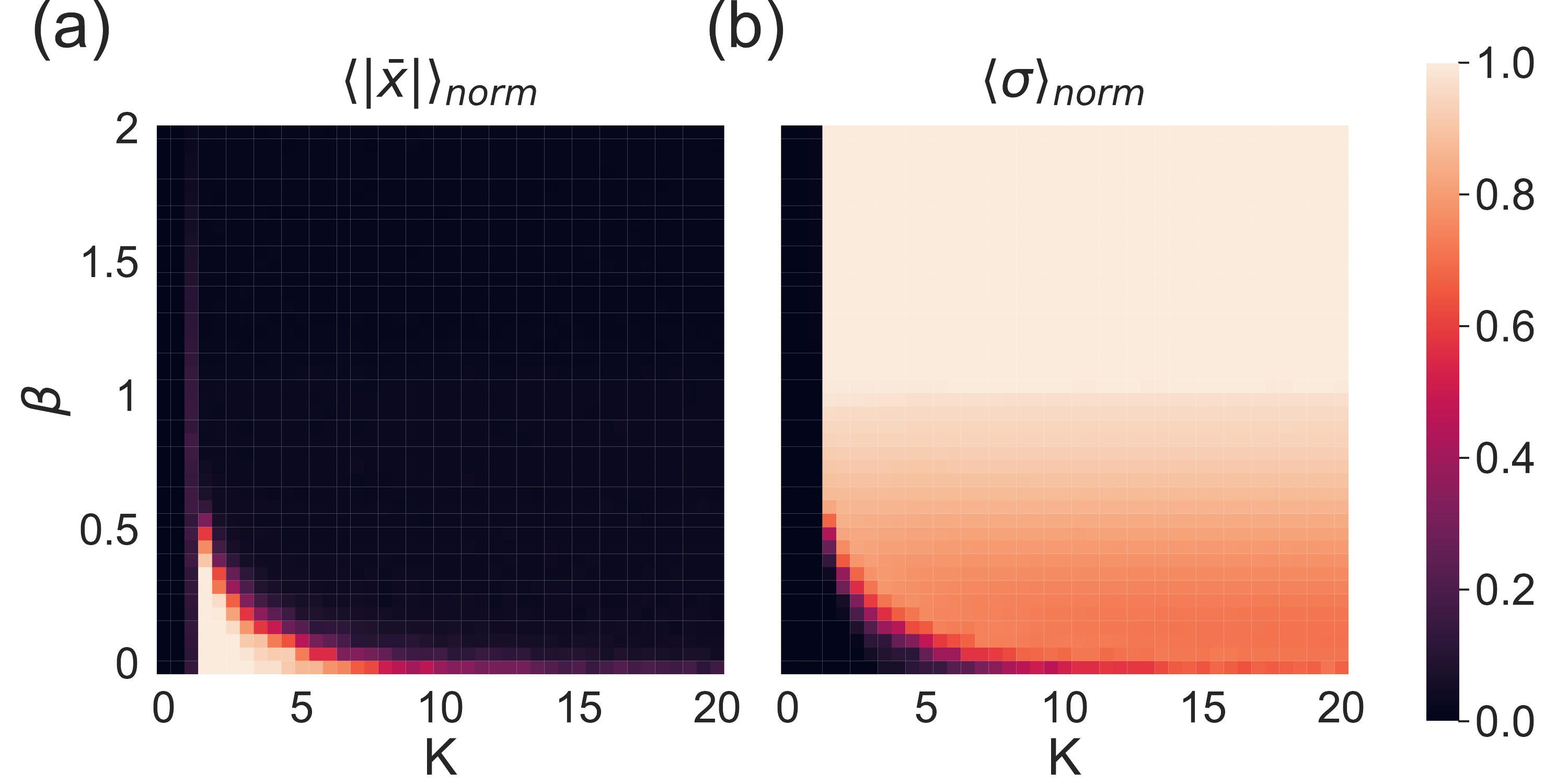}
    \caption{\label{fig:ER_k=4} (a) Average mean opinion $\langle |\bar{x}|\rangle_{norm}$ and (b) standard deviation $\langle \sigma\rangle_{norm}$ obtained from 100 independent realizations for each combination of parameters $(K, \beta)$ (color code). Lighter colors correspond to higher values and vice versa. In each realization, initial opinions are randomly chosen on the interval $[-x_{\text{max}}, x_{\text{max}}]$. The underlying structure of contacts throughout the figure is an Erd\"os-R\'enyi network of $N=9824$ nodes with mean degree $\langle k\rangle=4$.}
\end{figure}

Figs.~\ref{fig:ER}(a) and (b) show the phase diagrams obtained for an Erd\"os-R\'enyi network of $N=10^4$ nodes and mean degree $\langle k\rangle=10$. The process for obtaining these results is as follows: first, we perform 100 independent simulations for each combination of parameters $(K,\beta)$. Then, we compute the absolute value of the mean opinion $|\bar{x}|$ and standard deviation $\sigma$ of each of the generated opinion distributions, and average those results obtaining $\langle |\bar{x}|\rangle$ and $\langle \sigma \rangle$. Note that, as the range of opinions that an agent can take depends on $K$, results are divided by $x_{\max}$ so that they lie on the range $[0,1]$. In general, we observe the three phases predicted: for $K<1$ both measures are 0, resulting in neutral consensus states. For $K>1$ and $\beta>1$, the mean opinion is 0, and the standard deviation is maximum, corresponding to polarized states. Finally, for a certain range of parameters in the region $K>1$ and $\beta<1$, mean opinion reaches its maximum value, and the standard deviation remains near 0. Fig.~\ref{fig:ER}(c) shows some example trajectories of the equilibrium configurations previously mentioned, taken from the marked combination of parameters in Fig.~\ref{fig:ER}(a) and (b), and Fig.~\ref{fig:ER} shows the final opinion distributions in each of those cases.

An additional transition appears at $\beta_c(K)<1$. This transition splits the radicalized phase into two: for $\beta<\beta_c(K)$, configurations end up in radicalized states, while for $\beta_c(K)\leq \beta < 1$, some configurations end up being radicalized, while others reach metastable polarization states. For high enough $K$ and $\beta$, behavior becomes independent of $K$. These results prove that polarization is indeed reachable in networked populations, and almost always present and stable for $\beta>1$.

\begin{figure}[t]
    \centering
    \includegraphics[width=.5\textwidth]{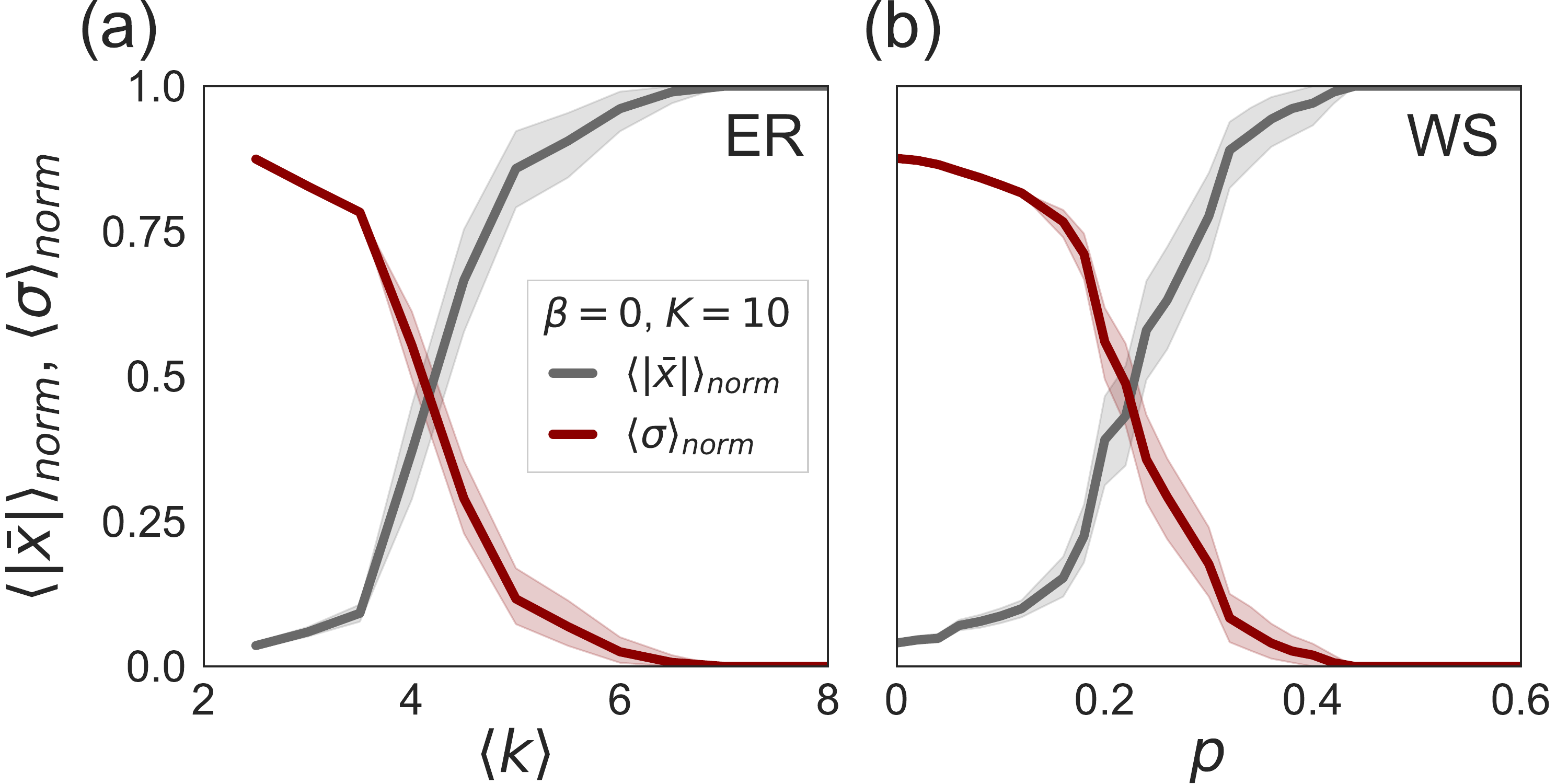}
    \caption{\label{fig:degree_rewiring} Average mean opinion $\langle|\bar{x}|\rangle_{norm}$ (gray) and standard deviation $\langle \sigma\rangle_{norm}$ (red) obtained from 100 independent simulations. Results are shown with 95\% confidence intervals (shadowed regions). $K=10$ and $\beta=0$ throughout the figure. (a) Results for an Erd\"os-R\'enyi network of $N=10^4$ nodes and varying mean degree $\langle k \rangle$. (b) Results for a Watts-Strogatz network of $N=10^4$ nodes and $k=10$, as a function of the rewiring probability $p$. The degrees of all nodes remain fixed in the rewiring process.}
\end{figure}

Fig.~\ref{fig:ER} reveals the existence of a complex interplay between the homophily mechanism and the structure of contacts in the emergence of polarization. To explore this, in Fig.~\ref{fig:ER_k=4} we reproduce the same results as in Fig.~\ref{fig:ER} using an Erd\"os-R\'enyi network of $N=9824$ nodes and mean degree $\langle k\rangle=4$. It can be seen that the transition at $\beta_c(K)<1$ is heavily dependent on the network degree, as is the probability of finding polarized states. In particular, more limited information horizons result in an increased polarization. Moreover, it can be seen in Fig.~\ref{fig:ER_k=4} that polarization can show up when $\beta=0$ (that is, in absence of homophily) from the network structure alone if the necessary features are present. To further investigate this phenomenon, we repeat our analysis by fixing $K=10$ and studying the evolution of both order parameters as we increase the average degree of the underlying ER network. Fig.~\ref{fig:degree_rewiring}(a) shows that lower degrees give rise to polarization, which quickly decays for increasing connectivity as the information horizon grows. Therefore, if connectivity is low enough, structural communities can breed local radicalization, and global polarization can be formed. The former phenomenon can also be driven by other network features determining the global communication in the network, such as the existence of structural correlations. To illustrate it, we consider a Watts-Strogatz graph in which we fix the degree $k=10$ and vary the correlations with the rewiring probability $p$, and show in Fig.~\ref{fig:degree_rewiring}(b) that this network can breed polarization provided a sufficiently small $p$. Taken together, these results highlight the importance of a limited information horizon on the rise of polarization. Note that this phenomenon cannot be characterized in models relying on temporal networks, in which network structure was either not present or evolved as a product of the polarization dynamics.

\begin{figure}[t]
    \centering
    \includegraphics[width=.5\textwidth]{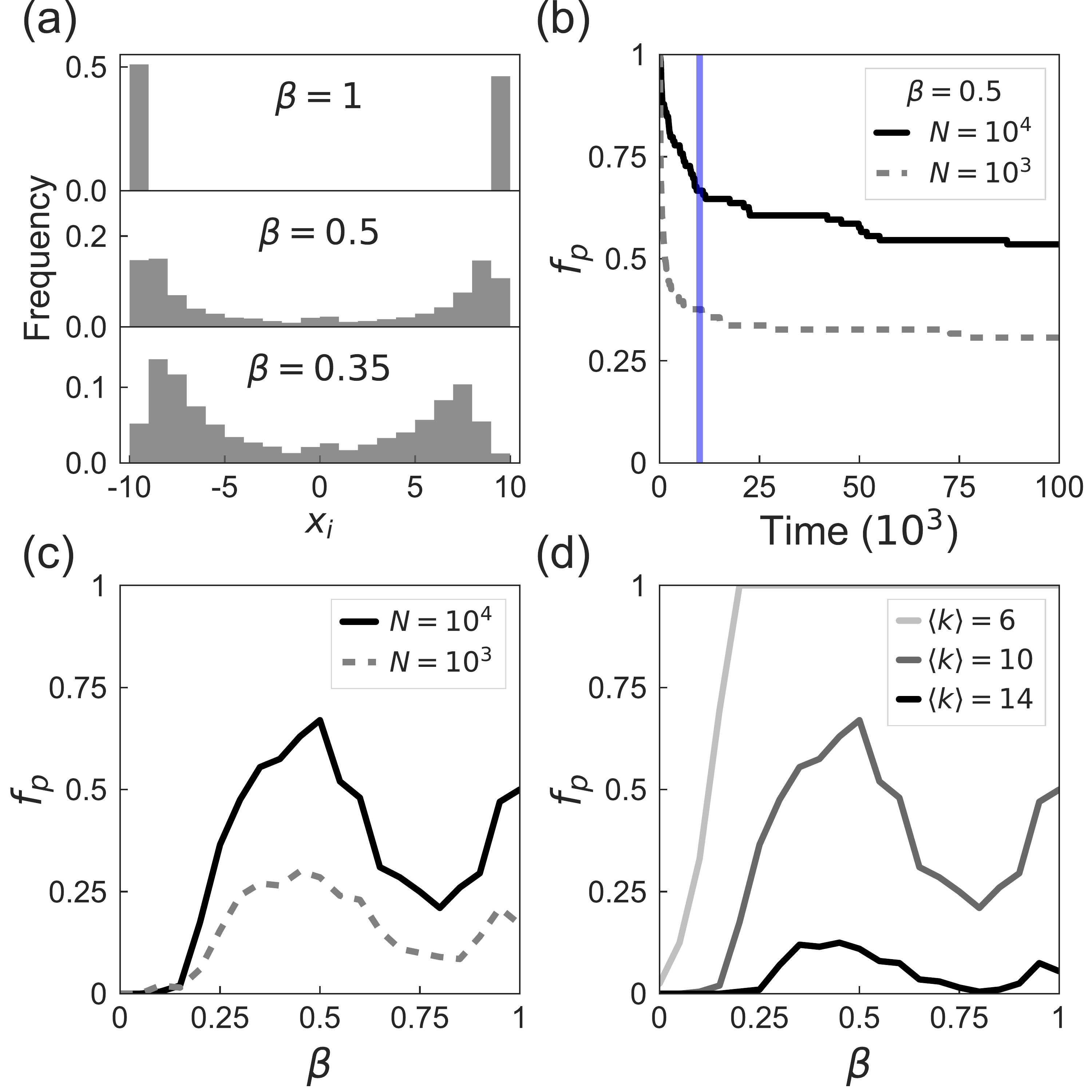} 
    \caption{\label{fig:polarization} (a) Opinion histograms of metastable polarized configurations obtained for different values of $\beta$. (b) Fraction of polarized configurations as a function of time for $\beta=0.5$, for an ER network of $\langle k\rangle=10$, and $10^4$ nodes (black line) or $10^3$ nodes (gray dashed line). To perform the study we select 100 configurations considered to be polarized at a time $t=10^2$ ($10^3$ steps), and simulate for a total of $10^5$ time units, corresponding to $10^6$ simulation steps. A configuration is considered to be polarized if the absolute value of its mean opinion $|\bar{x}|$ is bigger than the standard deviation $\sigma$. The blue vertical line marks the time chosen to establish our criterion of polarization in the rest of the figure, which is $10^5$ steps corresponding to $10^4$ time units. (c) Fraction of polarized configurations as a function of $\beta$ for different network sizes. Results correspond to an ER network of $\langle k\rangle=10$, comprised of $10^4$ nodes (black line), and $10^3$ nodes (gray dashed line). (d) Fraction of polarized configurations as a function of $\beta$ for different mean degrees. Results correspond to an ER network of $10^4$ nodes and $\langle k\rangle=6$ (light gray), $\langle k\rangle=10$ (gray), and $\langle k\rangle=14$ (black). In panels (c) and (d) 200 independent simulations are performed for each value of $\beta$. We fix $K=10$ throughout the figure.}
\end{figure}

Now we analyze the polarized opinions distributions generated by our model. In Fig.~\ref{fig:polarization}(a), we show how these configurations are strongly shaped by the homophily mechanism governed by the parameter $\beta$. Specifically, when $\beta$ is low, agents' opinions may not reach their maximum theoretical value, but instead remain in an intermediate standpoint. This comes in stark contrast to what can be found in previous models over static networks~\cite{meng2018networks}, in which agents tend to reach a finite limited set of final opinions. The position of the distributions' peaks depends on $\beta$, reaching the extreme value $x_{\text{max}}$ for $\beta \gtrsim 0.5$. By increasing $\beta$, extreme opinions become the most common and there is a depletion of agents holding intermediate positions until all agents reach the maximum theoretical opinion for $\beta \gtrsim 1$, falling in line with the stability analysis performed before. 

The metastable states occurring for $\beta_c(K)<\beta<1$ can reach very long lifetimes in some cases. To show this, we represent in Fig.~\ref{fig:polarization}(b) the fraction of configurations considered to be polarized at $t=100$, i.e. those for which $|\bar{x}|<\sigma$, which remain polarized at any given time $t$. There we check that, for these $\beta$ values, polarization quickly decays in the early stages of the evolution, but a sizeable amount of trajectories maintain polarized configurations for a very long time. Interestingly, this fraction increases with the size of the system, as a higher number of individuals makes heterogeneous local environments more likely to appear.

Fixing the simulation time to $t=10^4$ time units, we now study how the fraction of polarized configurations also depends on $\beta$, on the population size $N$, and the degree distribution $P(k)$. Restricting ourselves to $\beta<1$, we see in Fig.~\ref{fig:polarization}(c) that bigger networks always benefit polarization due to the larger space available for the formation of opposite, radicalized clusters. Furthermore, Fig.~\ref{fig:polarization}(d) shows that lower degrees dramatically increase the probability of reaching polarized states by limiting the information horizons of the agents, who become increasingly isolated. Finally, in the former figures, we can observe that, whenever the average connectivity $\langle k \rangle$ is large enough, polarization is the most probable around $\beta=0.5$ regardless of the network size and degree, corresponding to the value in which, as we stated before, the opinion histograms' peaks reach the maximum theoretical opinion values depleting intermediate positions.

A plausible explanation for this is that the network structure could lose its relevance for increasing $\beta$ due to this depletion process, as each agent would behave as if she was embedded in a fully connected network comprised of fully radicalized agents given by her environment. Following the argument made for fully connected networks, small perturbations of radicalized agents in proto-polarized configurations could provoke their de-radicalization and thus, effectively prevent the formation of polarized states. In turn, when most agents hold intermediate positions, the assumptions of the previous study no longer hold, agents' environments get richer, and the network structure could breed polarization, i.e.~generate equilibrium configurations for intermediate values of the agent's opinions.

\subsection{Mixing in polarized configurations}

Regardless of the network structure and features, the existence of stable contact patterns imply that agents usually become exposed to opposing views from disagreeing neighbors, forming cross-cutting ties that hinder the rise of echo chambers. This kind of ties is widespread on real-world societies, in which daily-life interactions can bring divergent views close together~\cite{mutz2002crosscutting, huckfeldt2004ambivalence, mutz2006workplace, goel2010perceived, bakshy2015facebook}. To measure the amount of cross-cutting interactions of a given configuration, we define the exposure $\mathcal{E}_i$ of agent $i$ as the fraction of neighbors that hold opinions of opposing sign with respect to the agents':
\begin{equation}
    \mathcal{E}_i = \frac{1}{k_i}\sum_{j=1}^N A_{ij}(1-\delta_{\sigma_i,\sigma_j}) \; ,
\end{equation} 

\noindent where $\sigma_i$ indicates the sign of agent $i$'s opinion. In the case of a randomly distributed population evenly split between both opinions, the expected average exposure is $\bar{\mathcal{E}}=0.5$.

These links do not necessarily imply an interaction capable of greatly disturbing an agents' opinion because of the importance that is given to each neighbor, determined by the weights $w_{ij}$. To measure the influence of cross-cutting ties over agent $i$, we define her attention $\mathcal{A}_i$ as the sum of her cross-cutting weights:
\begin{equation}
    \mathcal{A}_i = \sum_{j=1}^N A_{ij}w_{ij} (1-\delta_{\sigma_i,\sigma_j})\; .
\end{equation}

\begin{figure}[t]
    \centering
    \includegraphics[width=.5\textwidth]{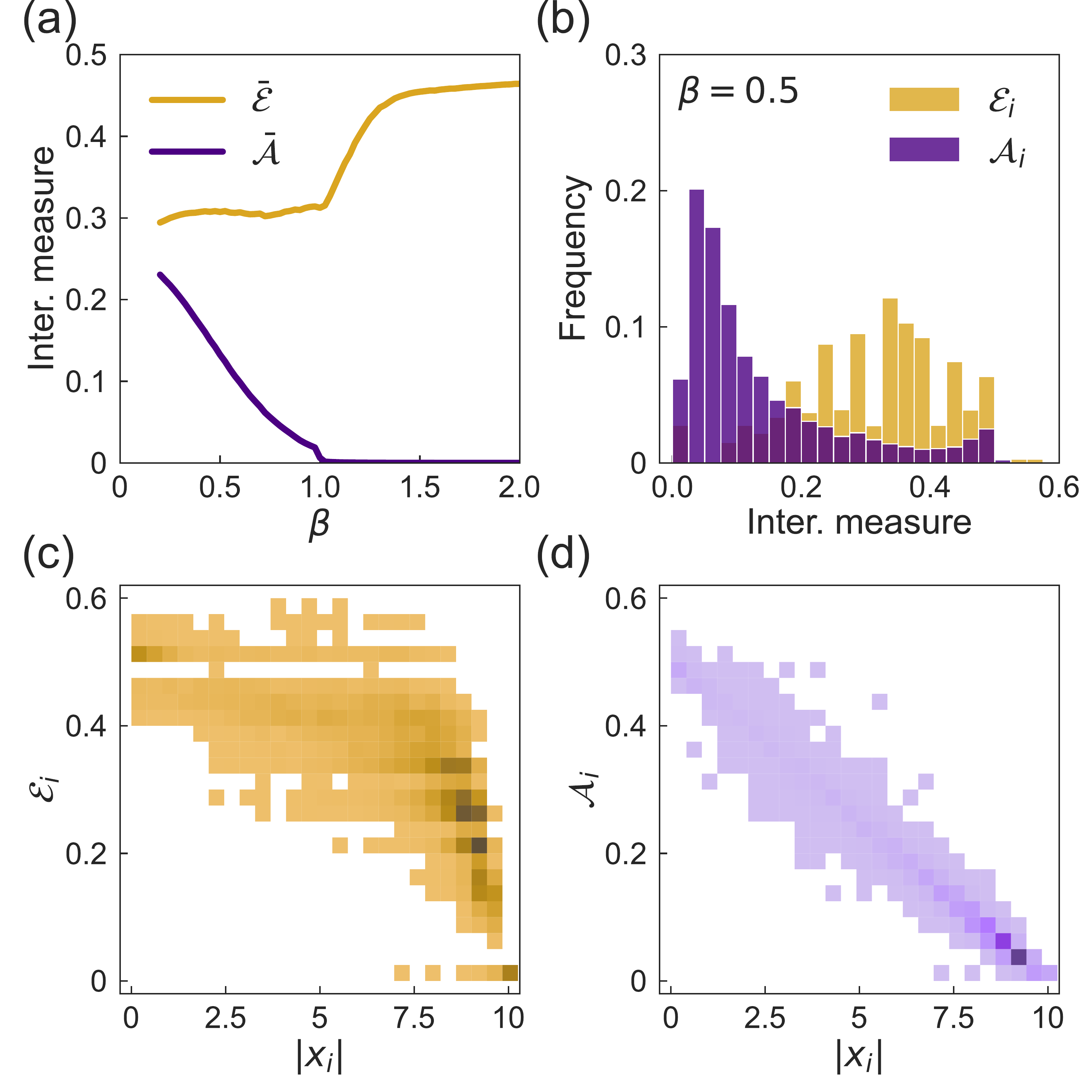}
    \caption{\label{fig:ER_inter_measures} (a) Average exposure (orange) and attention (purple) obtained from 100 independent simulations that reach polarization for each value of $\beta$. (b,c,d) Exposures and attentions of the polarized configuration shown Fig.~\ref{fig:polarization}(a) with $\beta=0.5$. (b) Frequency of exposures and attentions. (c) Two-dimensional histogram of individual exposure as a function of the absolute value of the agent's opinion. (d)  Two-dimensional histogram of individual attention as a function of the absolute value of the agent's opinion. In all the panels, we fix $K=10$ and consider an Erd\"os-R\'enyi network of $N=10^4$ nodes and $\langle k\rangle=10$.}
\end{figure}

\noindent As shown in Fig.~\ref{fig:ER_inter_measures}(a), the average exposure $\bar{\mathcal{E}}$ is significantly smaller than $0.5$ when $\beta<1$, indicating an aggregation of like-minded individuals in homophilic relationships. These opinion segregation phenomena can bear profound consequences for the propagation of diseases~\cite{carballosa2021epidemics, burgio2021tracing, burgio2022homophily} or misinformation~\cite{mocanu2015attention, bessi2015collective}. However, for $\beta>1$, $\bar{\mathcal{E}}$ increases monotonically towards the expected result of a well-mixed population. On the contrary, the average attention $\bar{\mathcal{A}}$ is a monotonically decreasing function of $\beta$, vanishing around $\beta=1$. Nevertheless, these results show that communication between disagreeing peers is still present even in polarized configurations, mimicking the behavior observed in real-world situations~\cite{huckfeldt2004ambivalence}.

As expected, higher homophily leads to less communication between disagreeing agents. The dependence of $\bar{\mathcal{E}}$ with $\beta$ can be therefore explained attending to $\bar{\mathcal{A}}$: smaller values of $\beta$ sustain information exchange in cross-cutting links, holding the agents in intermediate positions thanks to the influence of disagreeing neighbors. This, in turn, allows for the formation of bigger clusters of the same opinion as originally opposing individuals maintain some degree of information exchange, converging eventually. Higher values of $\beta$ promptly interrupt the communication between diverging neighbors, rendering the environment's influence negligible and giving rise to nearly well-mixed populations. Note that these results do not depend on the choice of $K$.

Focusing on the local environment of the individuals, $\mathcal{E}_i$ span a wide range of values regardless of the agents' opinions, as seen in Fig.~\ref{fig:ER_inter_measures}(c). However, it is clear that smaller $\mathcal{E}_i$ are strongly linked to intense radicalization, even though higher $\mathcal{E}_i$ do not guarantee milder views. In fact, Fig.~\ref{fig:ER_inter_measures}(d) shows that the main driver of radicalization is the attention $\mathcal{A}_i$ that an agent bestows on disagreeing neighbors: higher $\mathcal{A}_i$ is tantamount to weaker radicalization. Nevertheless, even for small values of $\beta$ most agents become isolated (albeit not completely disconnected) from opposing views, as can be seen in Fig.~\ref{fig:ER_inter_measures}(b). These results show that access to opposing opinions can prevent the polarization process as long as agents remain willing to consider other points of view~\cite{bail2018exposure, levy2021media}.

The linear relationship between opinion and attention observed in Fig.~\ref{fig:ER_inter_measures}(d) can be easily obtained from Eq.~\ref{eq:xChange}. Assuming $K$ high enough, the available opinion range will be $x_i \in (-K, K)$, and thus, $x_i$ will reach higher absolute opinion values. Then, $\tanh(x_i) \simeq \pm 1 $ for most agents, and:
\begin{equation}
    \dot{x}_i\simeq - x_i + K \sum_{j=1}^N A_{ij} w_{ij} \, \text{sgn}(x_i) = -x_i + K (w_{i,+} - w_{i,-}) \; ,
\end{equation}

\noindent where $w_{i,+}$ ($w_{i,-}$) represents the sum of weights with agents of positive (negative) opinions. Recalling the normalization of weights, $\sum_{j=1}^{N} A_{ij} w_{ij} = 1$, we have that $w_{i,\pm} = 1-w_{i,\mp}$, and then, in the equilibrium:
\begin{equation}
    x_i = \pm K(1-2w_{i,\mp}) \; ,
\end{equation}

\noindent unveiling the linear relationship. Following this equation, it is clear that, if $w_{i,+}>0.5$ ($w_{i,-}>0.5$), then $x_i>0$ ($x_i<0$), showing that agents bestow a greater amount of attention to like minded neighbors. However, this relationship only holds for $x_j$ high enough, so that $\tanh(x_j)\simeq \pm 1$. If this is not true, cross-cutting weights can reach values greater than 0.5 (as seen in Fig.~\ref{fig:ER_inter_measures}(d)).

\section{Experimental Data}
\label{sec:exp}

Finally, we apply our model to characterize the degree of polarization in the society with respect to different topics. For this purpose, we extract the opinions' distributions about traditionally polarized issues from the American National Election Study (ANES) of 2016~\cite{ANES2016}. The ANES analyzes the voting behavior and opinions of US society during presidential elections, by performing nation-wide surveys about multiple topics before and after the electoral process. We consider the topics corresponding to the following questions:

\begin{itemize}
    \item \textit{2010 health care law}: Do you favor, oppose, or neither favor nor oppose the health care reform law passed in 2010? This law requires all Americans to buy health insurance and requires health insurance companies to accept everyone. (code: V161114x)

    \item \textit{Voting as duty or choice}: Different people feel differently about voting. For some, voting is a duty - they feel they should vote in every election no matter how they feel about the candidates and parties. For others voting is a choice - they feel free to vote or not to vote, depending on how they feel about the candidates and parties. For you personally, is voting mainly a duty, mainly a choice, or neither a duty nor a choice? (code: V161151x)

    \item \textit{End birthright citizenship}: Some people have proposed that the U.S. Constitution should be changed so that the children of unauthorized immigrants do not automatically get citizenship if they are born in this country. Do you favor, oppose, or neither favor nor oppose this proposal? (code: V161194x)

    \item \textit{Wall with Mexico}: Do you favor, oppose, or neither favor nor oppose building a wall on the U.S. border with Mexico? (code: V161196x)

    \item \textit{Send troops to fight ISIS}: Do you favor, oppose, or neither favor nor oppose the U.S. sending ground troops to fight Islamic militants, such as ISIS, in Iraq and Syria? (code: V161213x)

    \item \textit{Allow to refuse service to same-sex couples}: Do you think business owners who provide wedding-related services should be allowed to refuse services to same-sex couples if same-sex marriage violates their religious beliefs, or do you think business owners should be required to provide services regardless of a couple’s sexual orientation? (code: V161227x)

    \item \textit{Transgender bathroom use}: Should transgender people – that is, people who identify themselves as the sex or gender different from the one they were born as – have to use the bathrooms of the gender they were born as, or should they be allowed to use the bathrooms of their identified gender? (code: V161228x)

    \item \textit{Free-trade agreements}: Do you favor, oppose, or neither favor nor oppose the U.S. making free trade agreements with other countries? (code: V162176x)

    \item \textit{Government spending for healthcare}: Do you favor an increase, decrease, or no change in government spending to help people pay for health insurance when they cannot pay for it all themselves? (code: V162193x)

    \item \textit{Torture for terrorists}: Do you favor, oppose, or neither favor nor oppose the U.S. government torturing people who are suspected of being terrorists, to try to get information? (code: V162295x)
    
\end{itemize}

The chosen questions correspond to different periods of surveying, before and after the election. In particular, issues ``Free-trade agreements", ``Government spending for healthcare", and ``Torture for terrorists" correspond to the post-election survey, and the rest, to the pre-election survey. Populations for both surveys vary slightly, and thus, different weights must be used to extrapolate the results to the global US population. Weights for the pre-election survey correspond to code V160201 of the data, and for the post-election survey, to code V160202.

Opinion histograms are normalized removing all the missing data, comprised of answers like ``don't know", ``refused to answer", survey errors or inexistence of survey data. Note that ``don't know" is considered to be different from ``Neither favor nor oppose" or other similar answers equidistant from the two extreme opinions.

\subsection{Comparison with the model}

Our aim is to find the values of $\beta$ and $K$ that generate the most similar opinion distributions to those obtained in surveys for polarized topics. To do that, we generate multiple polarized configurations for different values of the parameters, and compare their opinion distributions with the experimental ones. We use the Jensen-Shannon distance~\cite{lin1991jensenShannon} in order to measure the similarity between both distributions, and find the parameter combination that minimize that distance.

\begin{equation}
    d_{JS} = \sqrt{\frac{D(p||m) + D(q||m)}{2}} \; ,
\end{equation}

\noindent where $D(p||m)$ represents the Kullback-Leibler divergence:

\begin{equation}
    D(p||m) = \sum_i p(i) \ln{\frac{p(i)}{m(i)}} \; ,
\end{equation}

\noindent and $m$ is the pointwise mean of $p$ and $q$. Note that $d_{JS}$ is not defined if any value of $p$ or $q$ is zero, but no polarized distribution generated by our model in the relevant region of the parameter space fulfills this condition. It is also necessary that both distributions have the same number of points or boxes to compare them (most issues are comprised of seven data points, except for the issues ``Allow to refuse service to same-sex couples" and ``Transgender bathroom use" that have six data points). Therefore, we perform the following process over the synthetic distributions: first, for every polarized configuration generated for a given value of $K$, we divide the opinion value of each agent by $x_{\text{max}} (K)$, so that the final opinions lie on the range $(-1,1)$. Then, in order to map the continuous opinion distributions generated by the model into discrete ones, agent's opinions are aggregated into six or seven equally spaced boxes to match the number of distinct choices recorded for each survey.

\begin{figure}[t!]
    \centering
    \includegraphics[width=.5\textwidth]{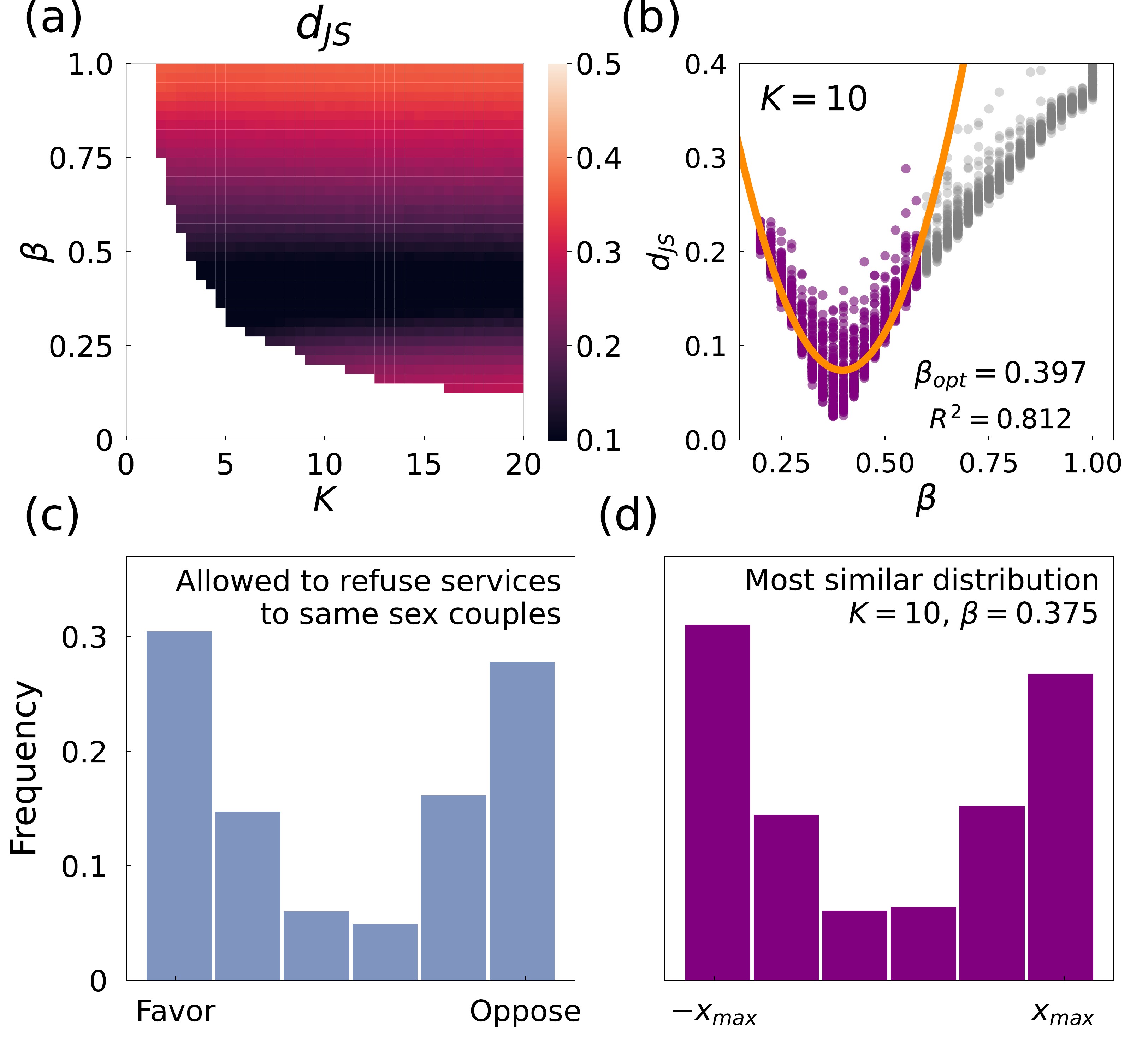}
    \caption{\label{fig:distances} Results for the issue ``Do you think business owners who provide wedding-related services should be allowed to refuse services to same-sex couples if same-sex marriage violates their religious beliefs?" of the ANES 2016 (codes V161227 and V161227a, aggregated in V161227x). (a) Mean Jensen-Shannon distance (color code) between the issue and polarized configurations generated by our model. 100 configurations are used for each combination of parameters $(K,\beta)$. Pairs of parameters whose probability of generating polarized configurations is less than $5\%$ are not considered. (b) Jensen-Shannon distances from all polarized configurations corresponding to $K=10$ to the distribution corresponding to the aforementioned issue. A second-degree function is fitted near the minimum to find $\beta_{\text{opt}}$ (orange line). In purple, distances used for the fit. In gray, distances not used for the fit. (c) Responses to the considered issue given by the ANES 2016 survey. (d) Most similar opinion distribution obtained in our simulations. In all the panels, we consider an Erd\"os-R\'enyi network of $N=10^4$ nodes and $\langle k\rangle=10$.}
\end{figure}

\begin{figure*}[t!]
    \centering
    \includegraphics[width=.74\textwidth]{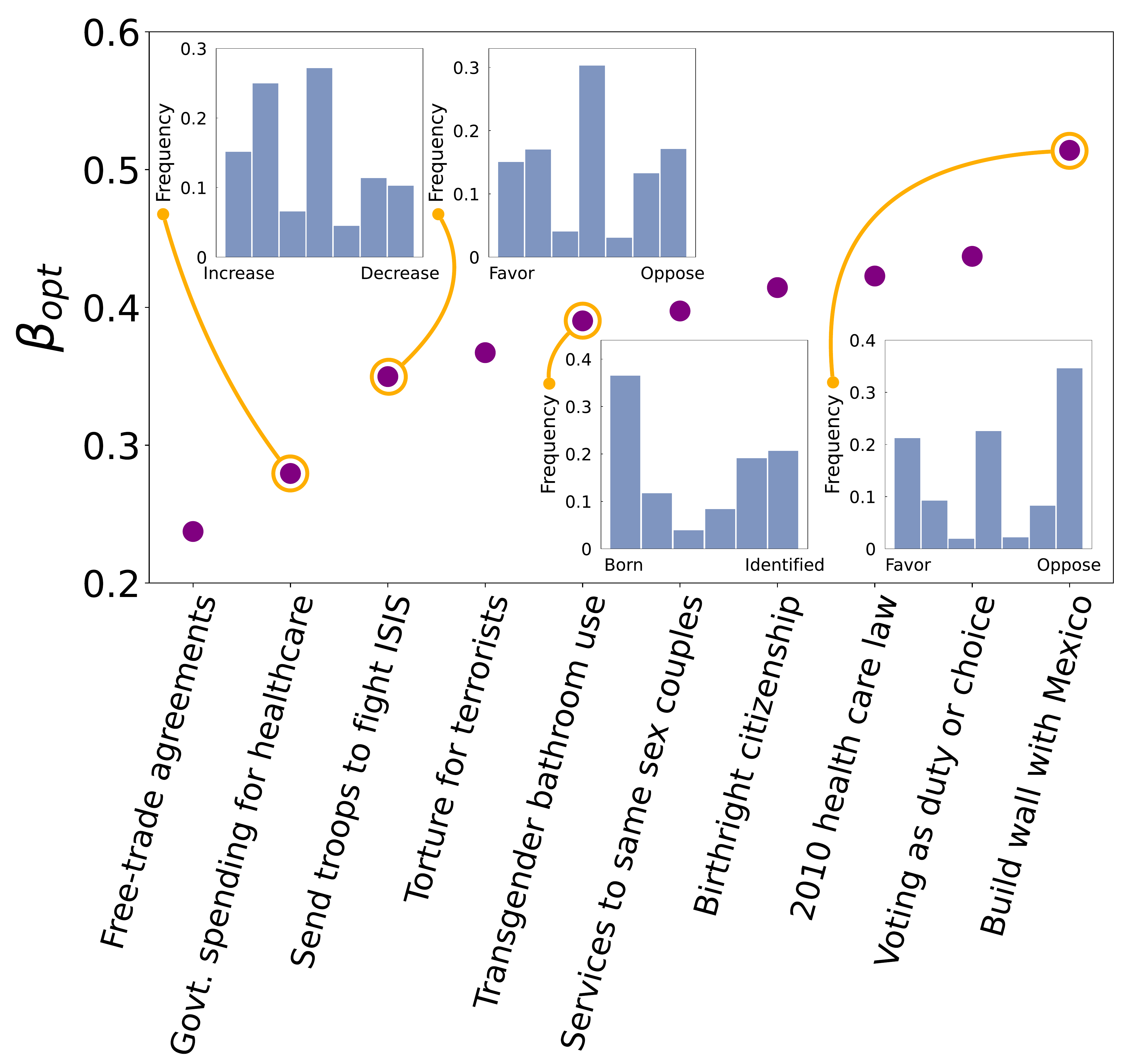}
    \caption{\label{fig:opt_betas} $\beta_{\text{opt}}$ estimated for multiple polarized issues of the ANES 2016 survey. Inset figures show some of their distributions. For the estimation process, we fix $K=10$ and consider an Erd\"os-R\'enyi network of $N=10^4$ nodes and $\langle k\rangle=10$. The errors of the estimations are obtained applying bootstrapping techniques, but they are too small to be visible.}
\end{figure*}

The shape of our opinion distributions is not dependent on $K$, which only determines the maximum reachable opinion. To show this, we generate 100 polarized configurations for multiple combinations of parameters $(K,\beta)$ in the range $K\in [0,20]$, $\beta\in[0,1]$ and compare them to the experimental one corresponding to the issue ``Refuse service to same sex couples" applying the method explained above. In Fig.~\ref{fig:distances}(a) we represent the mean $d_{JS}$ for each combination $(\beta,K)$, proving that there is no dependence with $K$ of the distances, and thus, we fix $K=10$ in the following process to estimate $\beta_{\text{opt}}$ for all issues.

In Fig.~\ref{fig:distances}(b) we show all distances between the given issue and the polarized distributions (gray and purple points) obtained for each value of $\beta$. As can be seen, there is a clear minimum that we can estimate by adjusting a second-degree function (orange line) in the vicinity of the smallest distance (purple points). The error of the adjustment is obtained applying bootstrapping techniques, but it is too small to be visible. For the sake of illustration, we represent in Fig.~\ref{fig:distances}(c) the experimental distribution for the selected issue, and in Fig.~\ref{fig:distances}(d), the most similar synthetic distribution, showing a great qualitative agreement among them.

Following our previous theoretical analysis of the model, $\beta_{opt}$ provides us with a way of sorting different issues by their degree of polarization. Results for all considered issues are shown in Fig.~\ref{fig:opt_betas}.  Issues with higher $\beta_{\text{opt}}$ show more extreme polarization, as agents for or against them tend to choose more extreme positions. Typical partisan topics, like the construction of a wall with Mexico, the 2010 health care law (commonly known as Obamacare), or the right of obtaining the citizenship by birth, follow this behavior. In contrast, issues that are not usually subject of mainstream political clash, or are perceived as extraneous by most of the population, like free trade agreements or the use of US troops to fight ISIS in the field, seem to arouse less exacerbating reactions. It is argued that racial resentment and fear of losing the dominant status of white working-class males were the main drivers of Donald Trump's victory in the US 2016 elections rather than economic issues~\cite{morgan2017elections, sides2017elections, abramowitz2019racial}. This falls in line with the opinion about free-trade agreements and government spending on healthcare being less polarized than other racial issues.

\section{Conclusions}
\label{sec:conc}

In conclusion, we have proposed a new opinion formation model applicable to static graphs, which introduces the homophily mechanism by making use of weights representing the attention an agent bestows on disagreeing neighbors. We have found that polarization is indeed reachable and stable in some cases while maintaining some amount of cross-cutting relationships, a typical situation of daily life social interaction, and that our opinion distributions reproduce qualitatively those of highly polarized issues observed in real-world surveys. Furthermore, we have shown that homophily is not essential for the rise of polarization if networks possess the appropriate features: namely low degree or high spatial correlations. These structural features favor the creation of limited information horizons, key in the stability of polarized states.

The phenomenon of polarization is very complex and has been discussed from multiple perspectives, with different definitions and measurements~\cite{lelkes2016measurements}. Following a criterion established in previous works from the complex systems perspective~\cite{hegselmann2002bounded,baumann2020}, we understand polarization as the existence of two opposing majority opinions following a bimodal-like distribution. This definition is also compatible with some perspectives coming from the political science field~\cite{dimaggio1996polarization, fiorina2008bimodal}, which considers bimodality as a necessary condition for polarization. The model also reproduces the phenomenon of "group polarization"~\cite{friedkin1999groupPolarization} coming from the field of social psychology, as whole populations can become increasingly radicalized towards one of the extreme opinions given the right circumstances. In this sense, our polarized distributions fit on the concept of "bi-polarization"~\cite{mas2013bipolarization}, as opposing groups remain polarized in antagonistic views.

There are several limitations to our approach. For the social dynamics, our model assumes that all individuals become engaged in the issue at hand with similar interest and willingness to discuss, ignores any predisposition of the agents towards a particular political orientation, and neglects other aspects like the existence of kinship ties, reputation, or professional expertise~\cite{small2013unimportant, offer2018demanding}. From a network perspective, for simplicity, all the results here discussed are obtained considering synthetic models to represent the contact structure of the population. Although we expect our results to be qualitatively robust with respect to the network choice, extending our analysis to real contact networks would allow disentangling the role of different network features such as degree heterogeneities, clustering or the existence of modular structures on the emergence and stability of polarized states. Nevertheless, our approach constitutes a step towards the study of opinion formation in contexts in which potentially opinion-changing interactions are bounded by other reasons apart from simple homophily, inherent to real-world social relationships.




\section*{Acknowledgments}
We acknowledge financial support from grant PID2020-113582GB-I00 funded by Spanish Ministerio de Ciencia e Innovaci\'on, and from Departamento de Industria e Innovaci\'on del Gobierno de Arag\'on y Fondo Social Europeo through projects no. E30\_23R (COMPHYS group) and E36\_23R (FENOL group). H.P.M. and F.B. acknowledge financial support from Gobierno de Arag\'on through doctoral fellowships. F.B. acknowledges financial support from grant PGC2018-094684-B-C22 funded by Spanish Ministerio de Ciencia e Innovaci\'on.

\bibliographystyle{elsarticle-bib_style} 
\bibliography{myrefs}

\begin{thebibliography}{10}
\expandafter\ifx\csname url\endcsname\relax
  \def\url#1{\texttt{#1}}\fi
\expandafter\ifx\csname urlprefix\endcsname\relax\def\urlprefix{URL }\fi
\expandafter\ifx\csname href\endcsname\relax
  \def\href#1#2{#2} \def\path#1{#1}\fi

\bibitem{baldassarri2007dynamics}
D.~Baldassarri, P.~Bearman, Dynamics of {{Political Polarization}}, American
  Sociological Review 72~(5) (2007) 784--811.
\newblock \href {https://doi.org/10.1177/000312240707200507}
  {\path{doi:10.1177/000312240707200507}}.

\bibitem{baldassarri2008constraint}
D.~Baldassarri, A.~Gelman, Partisans without {{Constraint}}: {{Political
  Polarization}} and {{Trends}} in {{American Public Opinion}}, American
  Journal of Sociology 114~(2) (2008) 408--446.
\newblock \href {https://doi.org/10.1086/590649} {\path{doi:10.1086/590649}}.

\bibitem{gentzkow2016polarization}
M.~Gentzkow, Polarization in 2016, Toulouse Network for Information Technology
  Whitepaper (2016) 1--23.

\bibitem{abramowitz2019racial}
A.~Abramowitz, J.~McCoy, United {{States}}: {{Racial Resentment}}, {{Negative
  Partisanship}}, and {{Polarization}} in {{Trump}}'s {{America}}, The ANNALS
  of the American Academy of Political and Social Science 681~(1) (2019)
  137--156.
\newblock \href {https://doi.org/10.1177/0002716218811309}
  {\path{doi:10.1177/0002716218811309}}.

\bibitem{wells2017conversation}
C.~Wells, K.~J. Cramer, M.~W. Wagner, G.~Alvarez, L.~A. Friedland, D.~V. Shah,
  L.~Bode, S.~Edgerly, I.~Gabay, C.~Franklin, When {We} {Stop} {Talking}
  {Politics}: {The} {Maintenance} and {Closing} of {Conversation} in
  {Contentious} {Times}, Journal of Communication 67~(1) (2017) 131--157.

\bibitem{mccoy2018democracy}
J.~McCoy, T.~Rahman, M.~Somer, Polarization and the {{Global Crisis}} of
  {{Democracy}}: {{Common Patterns}}, {{Dynamics}}, and {{Pernicious
  Consequences}} for {{Democratic Polities}}, American Behavioral Scientist
  62~(1) (2018) 16--42.
\newblock \href {https://doi.org/10.1177/0002764218759576}
  {\path{doi:10.1177/0002764218759576}}.

\bibitem{iyengar2018partisan}
S.~Iyengar, M.~Krupenkin, The {Strengthening} of {Partisan} {Affect}, Political
  Psychology 39~(S1) (2018) 201--218.
\newblock \href {https://doi.org/10.1111/pops.12487}
  {\path{doi:10.1111/pops.12487}}.

\bibitem{hersh2016physicians}
E.~D. Hersh, M.~N. Goldenberg, Democratic and {Republican} physicians provide
  different care on politicized health issues, Proceedings of the National
  Academy of Sciences 113~(42) (2016) 11811--11816, publisher: Proceedings of
  the National Academy of Sciences.
\newblock \href {https://doi.org/10.1073/pnas.1606609113}
  {\path{doi:10.1073/pnas.1606609113}}.

\bibitem{iyengar2015partisanship}
S.~Iyengar, S.~J. Westwood, Fear and {{Loathing}} across {{Party Lines}}: {{New
  Evidence}} on {{Group Polarization}}, American Journal of Political Science
  59~(3) (2015) 690--707.
\newblock \href {https://doi.org/10.1111/ajps.12152}
  {\path{doi:10.1111/ajps.12152}}.

\bibitem{huber2017dating}
G.~A. Huber, N.~Malhotra, Political {Homophily} in {Social} {Relationships}:
  {Evidence} from {Online} {Dating} {Behavior}, The Journal of Politics 79~(1)
  (2017) 269--283, publisher: The University of Chicago Press.
\newblock \href {https://doi.org/10.1086/687533} {\path{doi:10.1086/687533}}.

\bibitem{chen2018thanksgiving}
M.~K. Chen, R.~Rohla, The effect of partisanship and political advertising on
  close family ties, Science 360~(6392) (2018) 1020--1024.
\newblock \href {https://doi.org/10.1126/science.aaq1433}
  {\path{doi:10.1126/science.aaq1433}}.

\bibitem{farrell2012internet}
H.~Farrell, The {{Internet}}'s {{Consequences}} for {{Politics}}, Annual Review
  of Political Science 15~(1) (2012) 28.

\bibitem{allcott2020media}
H.~Allcott, L.~Braghieri, S.~Eichmeyer, M.~Gentzkow, The {{Welfare Effects}} of
  {{Social Media}}, American Economic Review 110~(3) (2020) 629--676.
\newblock \href {https://doi.org/10.1257/aer.20190658}
  {\path{doi:10.1257/aer.20190658}}.

\bibitem{schmidt2018vaccination}
A.~L. Schmidt, F.~Zollo, A.~Scala, C.~Betsch, W.~Quattrociocchi, Polarization
  of the vaccination debate on {{Facebook}}, Vaccine 36~(25) (2018) 3606--3612.
\newblock \href {https://doi.org/10.1016/j.vaccine.2018.05.040}
  {\path{doi:10.1016/j.vaccine.2018.05.040}}.

\bibitem{cinelli2021echo}
M.~Cinelli, G.~D.~F. Morales, A.~Galeazzi, W.~Quattrociocchi, M.~Starnini, The
  echo chamber effect on social media, Proceedings of the National Academy of
  Sciences 118~(9) (Mar. 2021).
\newblock \href {https://doi.org/10.1073/pnas.2023301118}
  {\path{doi:10.1073/pnas.2023301118}}.

\bibitem{levy2021media}
R.~Levy, Social {{Media}}, {{News Consumption}}, and {{Polarization}}:
  {{Evidence}} from a {{Field Experiment}}, American Economic Review 111~(3)
  (2021) 831--870.
\newblock \href {https://doi.org/10.1257/aer.20191777}
  {\path{doi:10.1257/aer.20191777}}.

\bibitem{garimella2021media}
K.~Garimella, T.~Smith, R.~Weiss, R.~West, Political {Polarization} in {Online}
  {News} {Consumption}, Proceedings of the International AAAI Conference on Web
  and Social Media 15 (2021) 152--162.
\newblock \href {https://doi.org/10.1609/icwsm.v15i1.18049}
  {\path{doi:10.1609/icwsm.v15i1.18049}}.

\bibitem{flaxman2016bubbles}
S.~Flaxman, S.~Goel, J.~M. Rao, Filter {{Bubbles}}, {{Echo Chambers}}, and
  {{Online News Consumption}}, Public Opinion Quarterly 80~(S1) (2016)
  298--320.
\newblock \href {https://doi.org/10.1093/poq/nfw006}
  {\path{doi:10.1093/poq/nfw006}}.

\bibitem{santos2021dynamical}
F.~P. Santos, Y.~Lelkes, S.~A. Levin, Link recommendation algorithms and
  dynamics of polarization in online social networks, Proceedings of the
  National Academy of Sciences 118~(50) (Dec. 2021).
\newblock \href {https://doi.org/10.1073/pnas.2102141118}
  {\path{doi:10.1073/pnas.2102141118}}.

\bibitem{sasahara2021unfollowing}
K.~Sasahara, W.~Chen, H.~Peng, G.~L. Ciampaglia, A.~Flammini, F.~Menczer,
  Social influence and unfollowing accelerate the emergence of echo chambers,
  Journal of Computational Social Science 4~(1) (2021) 381--402.
\newblock \href {https://doi.org/10.1007/s42001-020-00084-7}
  {\path{doi:10.1007/s42001-020-00084-7}}.

\bibitem{tokita2021cascade}
C.~K. Tokita, A.~M. Guess, C.~E. Tarnita, Polarized information ecosystems can
  reorganize social networks via information cascades, Proceedings of the
  National Academy of Sciences 118~(50) (Dec. 2021).
\newblock \href {https://doi.org/10.1073/pnas.2102147118}
  {\path{doi:10.1073/pnas.2102147118}}.

\bibitem{baumann2020}
F.~Baumann, P.~{Lorenz-Spreen}, I.~M. Sokolov, M.~Starnini, Modeling {{Echo
  Chambers}} and {{Polarization Dynamics}} in {{Social Networks}}, Physical
  Review Letters 124~(4) (2020) 048301.
\newblock \href {https://doi.org/10.1103/PhysRevLett.124.048301}
  {\path{doi:10.1103/PhysRevLett.124.048301}}.

\bibitem{baumann2021}
F.~Baumann, P.~{Lorenz-Spreen}, I.~M. Sokolov, M.~Starnini, Emergence of
  {{Polarized Ideological Opinions}} in {{Multidimensional Topic Spaces}},
  Physical Review X 11~(1) (2021) 011012.
\newblock \href {https://doi.org/10.1103/PhysRevX.11.011012}
  {\path{doi:10.1103/PhysRevX.11.011012}}.

\bibitem{dubois2018echo}
E.~Dubois, G.~Blank, The echo chamber is overstated: the moderating effect of
  political interest and diverse media, Information, Communication \& Society
  21~(5) (2018) 729--745.

\bibitem{guess2018echo}
A.~Guess, B.~Lyons, B.~Nyhan, J.~Reifler, Avoiding the echo chamber about echo
  chambers: {Why} selective exposure to like-minded political news is less
  prevalent than you think, Knight Foundation, 2018.

\bibitem{benkler2018propaganda}
Y.~Benkler, R.~Faris, H.~Roberts, Network {Propaganda}: {Manipulation},
  {Disinformation}, and {Radicalization} in American Politics, Oxford
  University Press, 2018.

\bibitem{zhuravskaya2020socialmedia}
E.~Zhuravskaya, M.~Petrova, R.~Enikolopov, Political {Effects} of the
  {Internet} and {Social} {Media}, Annual Review of Economics 12~(1) (2020)
  415--438.
\newblock \href {https://doi.org/10.1146/annurev-economics-081919-050239}
  {\path{doi:10.1146/annurev-economics-081919-050239}}.

\bibitem{barbera2014polarization}
P.~Barber\'a, How {Social} {Media} {Reduces} {Mass} {Political} {Polarization}.
  {Evidence} from {Germany}, {Spain}, and the {U}.{S}., Job Market Paper, New
  York University 46 (2014) 46.

\bibitem{boxell2017demographic}
L.~Boxell, M.~Gentzkow, J.~M. Shapiro, Greater {Internet} use is not associated
  with faster growth in political polarization among {US} demographic groups,
  Proceedings of the National Academy of Sciences 114~(40) (2017) 10612--10617.
\newblock \href {https://doi.org/10.1073/pnas.1706588114}
  {\path{doi:10.1073/pnas.1706588114}}.

\bibitem{huckfeldt2004ambivalence}
R.~Huckfeldt, J.~M. Mendez, T.~Osborn, Disagreement, {Ambivalence}, and
  {Engagement}: {The} {Political} {Consequences} of {Heterogeneous} {Networks},
  Political Psychology 25~(1) (2004) 65--95.
\newblock \href {https://doi.org/10.1111/j.1467-9221.2004.00357.x}
  {\path{doi:10.1111/j.1467-9221.2004.00357.x}}.

\bibitem{francisci2021reddit}
G.~De~Francisci~Morales, C.~Monti, M.~Starnini, No echo in the chambers of
  political interactions on {Reddit}, Scientific Reports 11~(1) (2021) 2818.
\newblock \href {https://doi.org/10.1038/s41598-021-81531-x}
  {\path{doi:10.1038/s41598-021-81531-x}}.

\bibitem{monti2023reddit}
C.~Monti, J.~D'Ignazi, M.~Starnini, G.~D.~F. Morales, Evidence of {Demographic}
  rather than {Ideological} {Segregation} in {News} {Discussion} on {Reddit}
  (Feb. 2023).
\newblock \href {https://doi.org/10.1145/3543507.3583468}
  {\path{doi:10.1145/3543507.3583468}}.

\bibitem{mutz2006workplace}
D.~C. Mutz, J.~J. Mondak, The {Workplace} as a {Context} for {Cross}-{Cutting}
  {Political} {Discourse}, The Journal of Politics 68~(1) (2006) 140--155.
\newblock \href {https://doi.org/10.1111/j.1468-2508.2006.00376.x}
  {\path{doi:10.1111/j.1468-2508.2006.00376.x}}.

\bibitem{goel2010perceived}
S.~Goel, W.~Mason, D.~J. Watts, Real and perceived attitude agreement in social
  networks, Journal of Personality and Social Psychology 99 (2010) 611--621.
\newblock \href {https://doi.org/10.1037/a0020697}
  {\path{doi:10.1037/a0020697}}.

\bibitem{bakshy2015facebook}
E.~Bakshy, S.~Messing, L.~A. Adamic, Exposure to ideologically diverse news and
  opinion on {Facebook}, Science 348~(6239) (2015) 1130--1132, publisher:
  American Association for the Advancement of Science.
\newblock \href {https://doi.org/10.1126/science.aaa1160}
  {\path{doi:10.1126/science.aaa1160}}.

\bibitem{offer2018demanding}
S.~Offer, C.~S. Fischer, Difficult {{People}}: {{Who Is Perceived}} to {{Be
  Demanding}} in {{Personal Networks}} and {{Why Are They There}}?, American
  Sociological Review 83~(1) (2018) 111--142.
\newblock \href {https://doi.org/10.1177/0003122417737951}
  {\path{doi:10.1177/0003122417737951}}.

\bibitem{deffuant2000bounded}
G.~Deffuant, D.~Neau, F.~Amblard, G.~Weisbuch, Mixing beliefs among interacting
  agents, Advances in Complex Systems 03~(01n04) (2000) 87--98, publisher:
  World Scientific Publishing Co.
\newblock \href {https://doi.org/10.1142/S0219525900000078}
  {\path{doi:10.1142/S0219525900000078}}.

\bibitem{hegselmann2002bounded}
R.~Hegselmann, U.~Krause, Opinion {Dynamics} and {Bounded} {Confidence}:
  {Models}, {Analysis} and {Simulation}, Journal of Artificial Societies and
  Social Simulation 5~(3) (2002).

\bibitem{meng2018networks}
X.~F. Meng, R.~A. Van~Gorder, M.~A. Porter, Opinion formation and distribution
  in a bounded-confidence model on various networks, Physical Review E 97~(2)
  (2018) 022312, publisher: American Physical Society.
\newblock \href {https://doi.org/10.1103/PhysRevE.97.022312}
  {\path{doi:10.1103/PhysRevE.97.022312}}.

\bibitem{lorenz2007review}
J.~Lorenz, Continuous opinion dynamics under bounded confidence: a survey,
  International Journal of Modern Physics C 18~(12) (2007) 1819--1838,
  publisher: World Scientific Publishing Co.
\newblock \href {https://doi.org/10.1142/S0129183107011789}
  {\path{doi:10.1142/S0129183107011789}}.

\bibitem{lorenz2008meanfield}
J.~Lorenz, Fixed points in models of continuous opinion dynamics under bounded
  confidence, arXiv:0806.1587 [physics] (Jun. 2008).
\newblock \href {https://doi.org/10.48550/arXiv.0806.1587}
  {\path{doi:10.48550/arXiv.0806.1587}}.

\bibitem{deffuant2002relative}
G.~Deffuant, F.~Amblard, G.~Weisbuch, T.~Faure, How can extremism prevail? a
  study based on the relative agreement interaction model, Journal of
  artificial societies and social simulation 5~(4) (2002).

\bibitem{deffuant2004smooth}
G.~Deffuant, F.~Amblard, G.~Weisbuch, Modelling {Group} {Opinion} {Shift} to
  {Extreme} : the {Smooth} {Bounded} {Confidence} {Model},
  arXiv:cond-mat/0410199 (Oct. 2004).
\newblock \href {https://doi.org/10.48550/arXiv.cond-mat/0410199}
  {\path{doi:10.48550/arXiv.cond-mat/0410199}}.

\bibitem{fiorina2008bimodal}
M.~P. Fiorina, S.~J. Abrams, Political {Polarization} in the {American}
  {Public}, Annual Review of Political Science 11:563-588 (2008).

\bibitem{ANES2016}
The american national election studies (www.electionstudies.org), these
  materials are based on work supported by the National Science Foundation
  under grant numbers SES 1444721, 2014-2017, the University of Michigan, and
  Stanford University.

\bibitem{mutz2002crosscutting}
D.~C. Mutz, The {Consequences} of {Cross}-{Cutting} {Networks} for {Political}
  {Participation}, American Journal of Political Science 46~(4) (2002)
  838--855, publisher: [Midwest Political Science Association, Wiley].
\newblock \href {https://doi.org/10.2307/3088437} {\path{doi:10.2307/3088437}}.

\bibitem{carballosa2021epidemics}
A.~Carballosa, M.~{Mussa-Juane}, A.~P. Mu{\~n}uzuri, Social {{Opinion
  Influence}} on {{Epidemic Scenarios}} (2021).

\bibitem{burgio2021tracing}
G.~Burgio, B.~Steinegger, G.~Rapisardi, A.~Arenas, The impact of homophily on
  digital proximity tracing, Physical Review Research 3~(3) (2021) 033128.
\newblock \href {http://arxiv.org/abs/2103.00635} {\path{arXiv:2103.00635}},
  \href {https://doi.org/10.1103/PhysRevResearch.3.033128}
  {\path{doi:10.1103/PhysRevResearch.3.033128}}.

\bibitem{burgio2022homophily}
G.~Burgio, B.~Steinegger, A.~Arenas, Homophily impacts the success of vaccine
  roll-outs, Communications Physics 5~(1) (2022) 70.
\newblock \href {https://doi.org/10.1038/s42005-022-00849-8}
  {\path{doi:10.1038/s42005-022-00849-8}}.

\bibitem{mocanu2015attention}
D.~Mocanu, L.~Rossi, Q.~Zhang, M.~Karsai, W.~Quattrociocchi, Collective
  attention in the age of (mis)information, Computers in Human Behavior 51
  (2015) 1198--1204.
\newblock \href {https://doi.org/10.1016/j.chb.2015.01.024}
  {\path{doi:10.1016/j.chb.2015.01.024}}.

\bibitem{bessi2015collective}
A.~Bessi, M.~Coletto, G.~A. Davidescu, A.~Scala, G.~Caldarelli,
  W.~Quattrociocchi, Science vs {{Conspiracy}}: {{Collective Narratives}} in
  the {{Age}} of {{Misinformation}}, PLOS ONE 10~(2) (2015) e0118093.
\newblock \href {https://doi.org/10.1371/journal.pone.0118093}
  {\path{doi:10.1371/journal.pone.0118093}}.

\bibitem{bail2018exposure}
C.~A. Bail, L.~P. Argyle, T.~W. Brown, J.~P. Bumpus, H.~Chen, M.~B.~F.
  Hunzaker, J.~Lee, M.~Mann, F.~Merhout, A.~Volfovsky, Exposure to opposing
  views on social media can increase political polarization, Proceedings of the
  National Academy of Sciences 115~(37) (2018) 9216--9221.
\newblock \href {https://doi.org/10.1073/pnas.1804840115}
  {\path{doi:10.1073/pnas.1804840115}}.

\bibitem{lin1991jensenShannon}
J.~Lin, Divergence measures based on the {{Shannon}} entropy, IEEE Transactions
  on Information Theory 37~(1) (1991) 145--151.
\newblock \href {https://doi.org/10.1109/18.61115}
  {\path{doi:10.1109/18.61115}}.

\bibitem{morgan2017elections}
S.~L. Morgan, J.~Lee, The {{White Working Class}} and {{Voter Turnout}} in
  {{U}}.{{S}}. {{Presidential Elections}}, 2004 to 2016, Sociological Science 4
  (2017) 656--685.
\newblock \href {https://doi.org/10.15195/v4.a27} {\path{doi:10.15195/v4.a27}}.

\bibitem{sides2017elections}
J.~Sides, M.~Tesler, L.~Vavreck, The 2016 {{U}}.{{S}}. {{Election}}: {{How
  Trump Lost}} and {{Won}}, Journal of Democracy 28~(2) (2017) 34--44.
\newblock \href {https://doi.org/10.1353/jod.2017.0022}
  {\path{doi:10.1353/jod.2017.0022}}.

\bibitem{lelkes2016measurements}
Y.~Lelkes, \href{https://doi.org/10.1093/poq/nfw005}{Mass {Polarization}:
  {Manifestations} and {Measurements}}, Public Opinion Quarterly 80~(S1) (2016)
  392--410.
\newblock \href {https://doi.org/10.1093/poq/nfw005}
  {\path{doi:10.1093/poq/nfw005}}.
\newline\urlprefix\url{https://doi.org/10.1093/poq/nfw005}

\bibitem{dimaggio1996polarization}
P.~DiMaggio, J.~Evans, B.~Bryson, Have {American}'s {Social} {Attitudes}
  {Become} {More} {Polarized}?, American Journal of Sociology 102~(3) (1996)
  690--755.
\newblock \href {https://doi.org/10.1086/230995} {\path{doi:10.1086/230995}}.

\bibitem{friedkin1999groupPolarization}
N.~E. Friedkin, Choice {Shift} and {Group} {Polarization}, American
  Sociological Review 64~(6) (1999) 856--875.
\newblock \href {https://doi.org/10.2307/2657407} {\path{doi:10.2307/2657407}}.

\bibitem{mas2013bipolarization}
M.~Mäs, A.~Flache, Differentiation without {Distancing}. {Explaining}
  {Bi}-{Polarization} of {Opinions} without {Negative} {Influence}, PLOS ONE
  8~(11) (2013) e74516.
\newblock \href {https://doi.org/10.1371/journal.pone.0074516}
  {\path{doi:10.1371/journal.pone.0074516}}.

\bibitem{small2013unimportant}
M.~L. Small, Weak ties and the core discussion network: {Why} people regularly
  discuss important matters with unimportant alters, Social Networks 35~(3)
  (2013) 470--483.
\newblock \href {https://doi.org/10.1016/j.socnet.2013.05.004}
  {\path{doi:10.1016/j.socnet.2013.05.004}}.

\end{thebibliography}





\end{document}